**Richard Smiraglia**
School of Information Studies, University of Wisconsin, Milwaukee, USA

**Andrea Scharnhorst**
Data Archiving and Networked Services, Royal Netherlands Academy of Art and Sciences, The Hague and Amsterdam, The Netherlands

**Almila Akdag Salah**
University of Amsterdam, Amsterdam, The Netherlands

**Cheng Gao**
Dalian, China

# UDC in Action

**Abstract:** The UDC is not only a classification language with a long history; it also presents a complex cognitive system worthy of the attention of complexity theory. The elements of the UDC: classes, auxiliaries, and operations are combined into symbolic strings, which in essence represent a complex networks of concepts.  This network forms a backbone of ordering of knowledge and at the same time allows expression of different perspectives on various products of human knowledge production. In this paper we look at UDC strings derived from the holdings of libraries. In particular we analyse the subject headings of holdings of the university library in Leuven, and an extraction of UDC numbers from the OCLC WorldCat. Comparing those sets with the Master Reference File, we look into the length of strings, the occurrence of different auxiliary signs, and the resulting connections between UDC classes. We apply methods and representations from complexity theory. Mapping out basic statistics on UDC classes as used in libraries from a point of view of complexity theory bears different benefits. Deploying its structure could serve as an overview and basic information for users among the nature and focus of specific collections. A closer view into combined UDC numbers reveals the complex nature of the UDC as an example for a knowledge ordering system, which deserves future exploration from a complexity theoretical perspective.

**Keywords:** complexity, visualization, statistical analysis, evolution, structure, classification

**1. Introduction**

Inherent in all knowledge domains is a certain systematization of knowledge about phenomena. But, as Kedrov (1975) expresses it, the need to classify all of the knowledge of humankind regularly emerges. This is the domain in which universal classifications, such as the Universal Decimal Classification (UDC), operate (Slavic, 2008a) (Slavic, 2008b). The design of classification systems used to order knowledge and retrieve information entails a wealth of contextual information about which concepts are thought to be most relevant, what is the envisioned relationship between them, and how best to express this in a formal way. Those system designs can be approached as though they were complex phenomena of human knowledge production. In this paper we apply some typical methods for studying complex systems based on quantitative information that can be gathered around them.

Classification research is the scientific field where traditionally form, structure, meaning, and increasingly also the evolution of classification systems (Tennis, 2007) are studied and implications for information retrieval are discussed (Tennis, 2006). Still in this field, quantitative methods – not to mention large-scale data analysis and accompanying visual explorations (Osinska, 2010; Börner, 2010) are not yet standard elements in the academic discourse. Digital representations of classification systems (see the Master Reference File of the UDC as one example) and their instantiations in collections of musea and libraries (see on-line catalogues and OAI-PHM based API's) make, enable, and call for computer-based methods of analysis. Moreover, quantitative indicators about the size, inner composition, growth and change of a classification system are important for the monitoring of the design of the system as well as of its application. So, it is no surprise that this kind of information is collected and analyzed, for instance, by the UDC editors in the process of the maintenance and revision of UDC (McIlwaine, 2007). However, the main focus of makers and users of the UDC, as of the makers of other classification systems, is not the scientific analysis of this system on a fundamental level. This might explain the lack of systematic quantitative studies about these systems. For outsiders, on the other hand, it is not always easy to have access to numeric information about the implementation of the UDC, and comparable systems.

In earlier studies the authors of this paper have used the published editions of the MRF of the UDC, a printed early version (manual data entered) and information from secondary sources as the Guideline editions to display, analyze and discuss structure and evolution of the UDC classes and the use of auxiliaries in the UDC (Scharnhorst et al. 2011; Akdag Salah et al. 2012a; Scharnhorst & Smiraglia 2012). In a comparative study of UDC and Wikipedia category structures (Akdag Salah et al. 2011; 2012b) undertaken by some of the authors it was evident that to some extent pears and apples were compared: the Wikipedia being a collection combined with an emergent category system, the UDC in its Master Reference File instantiation being a designed system only. This led the team to wish to also look into the actual use of the UDC.

This paper presents results from the study of two sets of UDC numbers retrieved from 'living systems'. Thanks to the courtesy of colleagues we received a sample of UDC numbers as used in the library of the *Katholieke Universiteit Leuven* (KU Leuven, retrieved May 2011); and an extraction of UDC numbers that were found in WorldCat (May 2011). We demonstrate insights based on statistical analysis and corresponding visuals to understand not only the envisioned but also the actual use of a classification system.

## 2. Datasets

### 2.1. OCLC

The OCLC dataset is an ASCII text file. The file has 9,055,623 records extracted from 214,596,487 bibliographic records using the "080" field in WorldCat. Each of these lines contains two strings (see table 1). The first column obviously represents an internal ID number. The second column contains records with UDC numbers, always behind the MARC subfield an indicator. Some lines in the file do not start with an 'a' tag. Each of the lines represents a book, which has been assigned a UDC number. Leaving out the non-UDC number records (those records in which there are no numbers after 'a' or there is no 'a' at all), there are 8,944,669 records in total. Of these records another 570,629 have been dismissed as non-UDC numbers. Eventually, we identified 8,374,040 lines containing UDC numbers; there were single numbers as well as composed numbers on which the further analysis is based.

**Table1:** *Original file of extracted numbers from OCLC WorldCat – first 10 lines*

| 16213477 | a621.315.2:678.742:621.395 |
|---|---|
| 17323477 | a614.25 |
| 17343477 | a618.177:616.697 |
| 36603477 | a82-31/-32 |
| 36613477 | a597 |
| 36613477 | a82-93 |
| 36663477 | a614.253.5.001 |
| 37253477 | a82-93 |
| 37253477 | a820(73)-31 |
| 37963477 | a577 |

### 2.2. Leuven

The set from Leuven is slightly different. It represents the UDC-classifications used in the LIBIS systems of the university library of Leuven. The original file has 95,544 lines. The first column contains a string with the structure $$8 UDC number $$a UDC heading $$9 language of the heading. The second says how often this UDC number is used in bib records in the library. Table 2 line 1 shows the UDC number "18" = an auxiliary number for time, and the headings representing entries for the nineteenth century. This auxiliary can be found twice in the bib records of the library.

**Table2:** *Original file UDC headings_LIBISnet - Leuven set – first 10 lines*

| $$8"18"$$a19e eeuw. Periode 1800-1899$$9dut | 2 |
|---|---|
| $$8(043)U1$$aDissertaties--U1$$9dut | 6 |



| | |
|---|---|
| $$8(043)U2$$aDissertaties--U2$$9dut | 3 |
| $$8(043)W1$$aDissertaties--W1$$9dut | 6 |
| $$8001 <03>$$aWetenschap en kennis--(algemeen)--Naslagwerken. Referentiewerken$$9dut | 3 |
| $$8001 <061>$$aWetenschap en kennis--(algemeen)--?<061>$$9dut | 1 |

Specific to the KU library is the use of UDC numbers for subject headings. A book record from their catalogue looks like this:

Title
Social structure and change : Finland and Poland : comparative perspective / Ed. by Erik Allardt and Wlodzimierz Wesolowski.
Publ. Year
1978.
Publisher
 Warszawa : Polish scientific publ.,
Editor(s)
 Allardt, Erik /  Wesołowski, Włodzimierz /
Physical details
391 p.
Subjects
 Social change. Sociale ontwikkeling. Sociale veranderingen. Modernisering. Evolutie .Sociale revolutie. Modernisme : 316.42
 Sociale structuur --(sociologie) --Finland : 316.3 <480>
 Sociale structuur --(sociologie) -- Polen : 316.3 <438>

This book record would belong to two entries in the file used here. Rather than representing books in UDC classification, this file represents the subject headings presently used, and how often they are used. For the sake of simplicity we decided in a first step to ignore the frequency of use (second column in Table 2), and analysed and aggregated the unique occurrences. Eventually, for Leuven 91,132 lines with UDC numbers were analysed.

Additional to these two 'real world' data sets we also used the Master Reference File from the UDC (edition 2008), called MRF in the following.

### 3. Data processing

Once the UDC numbers are extracted from the datasets, they need to be parsed in order to analyse them further. UDC numbers are strings, which contain numeric characters as well as other specific symbols. In the parsing attention must be paid to the sequence of numeric and non-numeric symbols. For example, the OCLC dataset contains the string:

**394.4 :[92(100+437) :329(437).15(091)+327.32(100)]**

First, **394.4** is one UDC class number standing for "Public ceremonial, coronations," and "**:**" is an auxiliary sign that stands for "simple relation." Square brackets are used for subgrouping. Everything within the **[….]** brackets is a unity. This unity starts with another UDC class number **92**, standing for "History or Biography". The () parentheses when starting with a non-zero numeric character point to a common auxiliary number of place. In our case **(100+437)** indicates "all countries in general 100" AND "Czech Republic 437". The next part of the unity in the square brackets starts again with UDC number **329**, standing for "Political parties" combined with an auxiliary of place, logically again the Czech Republic **(437)**. But this time the auxiliary of place is followed by an auxiliary of time **.015** standing for "second millennium CE," which is followed by another common auxiliary of form (0…). (For instance, **(091)** means manuscripts.) The last part in the unity starts with a "+" sign, the common auxiliary sign for Addition, pointing to the next UDC number in the string, 327.32 (Religious organizations) and extended with a common auxiliary of place again, namely (100) meaning "all countries in general." For an expert the UDC string above clearly points to a work dealing with

**"Public coronations in the History of political and religious influence on the Czech Republic in the 2nd millennium CE."[1]**

For the intended statistical analysis, we calculated the length of the strings, but also extracted information about which UDC class occurred in combination with which other class. We also differentiated between the different operations possible for combining UDC numbers. So, for the later analysis, the number above would deliver a tick in a matrix of row and columns standing for the UDC classes, between class 3 and class 3,9,3 taking account of the ":" operation of relation. It would also contribute a tick between class 3 and class 3 because of the "+" operation. But no operation would be counted for the (100+437) part, because in this case it would be an operation among common auxiliaries of place, and not among classes. This illustration also shows that operation based on parsing can easily go wrong if any symbols are ignored. At the beginning of our experiments we received quite a number of entries for operations that seemed to come from class 4. Knowing that class 4 is empty and not used in the current UDC, we went back to the parsing and found that all 4xx cases actually emerged from auxiliaries of place.

---

[1] Communication with the UDC Editor in Chief after submission illustrate the problem of complex UDC number in their application to describe works as well as in their interpretation by others than the original maker of the combination. It needs indepth familarity with the system to get the right interpretation: Aida Slavic writes "You misunderstood the following string 329(437).15 (091). This means " History of Communist party of Czechoslovakia". This is not surprising as the combination produced above is not typical … and has unnecessary complications and additions. The whole combination above describes the content to do with celebrations, parades and historical artifacts to do with flags, symbols, insignia, banners, persons speaking and participating on the celebrations/holidays in the history of Czech Communist party. The celebrations and public festivals themselves are related to the international movements of workers e.g. 1st May parades and 8th March parades. So instead of "public coronation" the number points to official celebratory days in socialist countries.

UDC numbers 'in the wild,' so to say, can also contain symbols other than those defined in the UDC. For instance <> brackets sometimes are used to add other notation or information to the UDC number. We ignored information that appeared in those brackets. The asterisk sign ""also can be used to introduce non-UDC notations, and so it was equally ignored in our parsing.

## 4. UDC in Action I – profile of collections

Based on the OCLC set and the Leuven set, we can now compare the actual use of the UDC in book cataloguing as well as in subject heading use together with the structure of the MRF file. We use MRF 2008 in the following. There is one more problem to address before we can turn to the distributions themselves. Both the OCLC and the Leuven set contain strings of simple and combined UDC numbers, which represent common auxiliary signs in conjunction with UDC numbers. The last can be understood as operations (similar to algebraic operations).

For the distribution of UDC numbers across the 10 main classes we applied two methods of counting. For the first method, we only used the UDC number at the beginning of the string. In the second method of counting, each UDC number (before and after an auxiliary sign) was counted. So, if we look at the arbitrary string: 123.456:213.465. From this string the first counting method would deliver one occurrence of class 1. With the second counting method both class 1 and class 2 would receive a tick of occurrence. Concerning the class distribution there is no big difference between the methods, so here we only display results based on counting method 1.

The MRF set of 2008 is with 54,980 records the smallest in absolute terms, followed by Leuven with 91,132 records. The OCLC set with 8,374,040 analyzed lines is three orders of magnitude larger.

We have argued earlier that the MRF to a certain extent reflects how changes in the sciences are perceived and accommodated in the editorial processes. The primary use of the UDC for academic libraries explains the relatively large size of class 6. In comparison the subject headings of Leuven are dominated by Social Sciences (3), followed by Religion (2), and Applied Sciences (6) in third place. In the OCLC dataset, Social Sciences (3) are leading, followed by Applied Sciences (6) and Language, Linguistics and Literature (8).

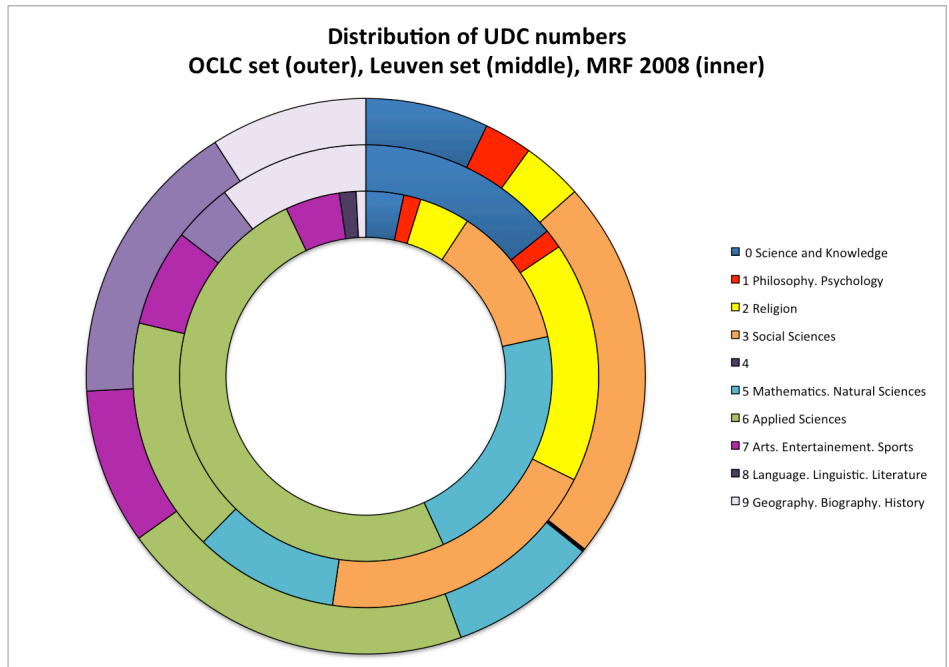

**Figure 1:** *Distribution of UDC numbers across classes, in design as well as in practical use.*

The actual use of UDC numbers mirrors the content of library collections (or an agglomeration of them as is found in OCLC). Not surpringsingly actual use deviates from the MRF, illustrating the collective diverse foci of collections.

## 5. UDC in Action II – analysis of the complexity of the UDC

### 5.1. Distribution of UDC strings

In the preceding section we already pointed to the complexity of the strings appearing in bibliographic records using UDC numbers, and the challenges for parsing. The UDC is a complex language that can capture the complex relationship between concepts rather than a static reference system in which concepts can be easily and uniquely placed. From the point of view of complexity, we took the length of a string as an indication of the complexity of a UDC number expressing the extent of recombination of concepts. The MRF file contains not only single UDC numbers, but different combinations of them leading to strings longer than just six tokens, which is the maximum length of a single UDC number.

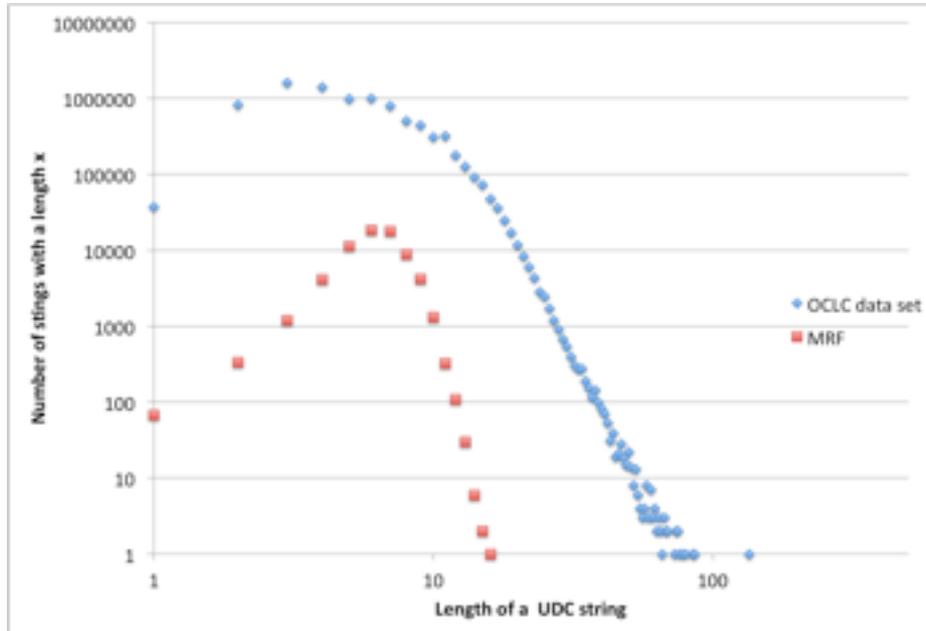

**Figure 1:**

How do we read the graphics in Figure 2, which is a typical representation in statistical physics called a distribution function? The first (red) data point represents all strings in the UDC MRF of size 1. There are 68 occurrences of strings of the length 1. Among them are the 9 main classes, but also other signs. Please do note that both x- and y-axe use a logarithmic scale. The distribution of string length in the MRF file peaks at length 6 (the length of single UDC numbers). But shorter or longer UDC numbers are not normally distributed (in a Gaussian curve) around this expected most frequent length. As in many other complex socio-technical systems (Scharnhorst 2003) the distribution function is skew, relatively long strings appear less frequently, the maximum string length in the MRF 2008 is 16. If we look into the same distribution for the OCLC data set, we immediately see that in actual use much more complex UDC numbers occur. They fall in the long tail of the distribution and can extend up to length 135. At the very least this distribution indicates that the UDC as a classification system shows features of a complex and non-linear system.

## 6. Weaving meaning from operations on classes – network views of the UDC

The UDC knows six operations ordered in three so-called Tables a, b and h.

- Table 1a contains + *Coordination. Addition (plus sign)* and */ Consecutive extension (oblique stroke sign)*. 1/5=1+2+3+4+5 / = range ….The second operation allows

- reducing redundancy when combining UDC numbers. For example, the operation 629.734/.735 is equivalent to 629.734+629.735. In general the Table1a auxiliary signs are used to extend subjects.
- Table 1b contains *: Simple relation (colon sign)*; *:: Order-fixing (double colon sign)*; *[] Subgrouping (square brackets)*. Different from Table 1a, Table 1b auxiliary signs are used rather to restrict subjects by defining relations among them.[2]
- Table 1h contains two signs which indicate non UDC notations: *Introduces non-UDC notation (asterisk)*; and *A/Z Direct alphabetical specification*.

In this paper we restrict the analysis to the operations "+" (plus), "/" (stroke) and ":" (colon).

To analyse which kind of operations, or in other words to discover which auxiliaries and auxiliary signs are responsible for the length or complexity of UDC numbers, we parsed the OCLC and the Leuven data set and constructed matrices around common auxiliary signs. As discussed previously in the section on data processing, the parsing is not always straightforward, and in the case of large data sets manual cleaning is out of the question. For control, we kept the class 4 (known to be empty) in the matrix. Cell values for class 4 indicate the level of noise and are marked in grey.

The sequence in which UDC classes are combined usually has meaning. So, all matrices are non-symmetrical and the related networks are directional. For example, the UDC number *022:11.203:042* is recorded in the matrix_colon table as a link between class 0 and class 1, but also in the same matrix_colon table as a link between class 1 and class 0. The combined number *022:11.203+11.204* contributes one tick to the cell {row class 0, column class 1} in the matrix_colon, and in the matrix_plus between row class 1 and column class 1. Combinations between auxiliaries are not taken into account. Table 3 in the appendix contains the matrices of auxiliary operations from the Leuven data set. For Leuven we only found the common auxiliaries "/" and ":" among classes. Table 4 contains the matrices of operations "+", "/" and ":" for the OCLC dataset.

The network visualization of these matrices is displayed in Figure 3 (Leuven) and Figures 4 (OCLC). For this we abandoned the class 4. The weight of links between classes is indicated by the width of the lines. The size of the nodes corresponds to the relative weighted degree of a node, and indicates the extent to which a class is linked up with other classes. We have not normalized the weight of links by the number of occurrence of a class in the dataset.

In the Leuven dataset, we found no occurrence of the use of the "+" Addition operation between classes. The Consecutive extension "/"can be found inside classes. This is indicated by the self-loops to one node. But the operation is also very popular in links between classes 0, 1, 2 and 3 (Fig 3b). The Simple relation ":" – equally often applied in total, is much more evenly used among all UDC classes (Fig 3a). In the OCLC dataset, we see the operation "+" (Addition) occur mostly in classes such as class 6, 3 and 8 (see Table 4). But, those classes are relatively large. So, one could also apply a normalization to the link weight based on the absolute occurrence of a class in a dataset. The Consecutive extension ("/") occurs one order of magnitude more frequently than "+" Addition. It is also clearly an operation that runs mostly within a class (Fig 4b). But the most popular operation among the three is "Simple relation." This operation (":") is most prominent among class 0 and 8 as can been seen both visually as examine the corresponding matrix (Fig 4a).

---

[2] See Outline of the UDC http://udcc.org/udcsummary/php/index.php?tag=---&lang=en

In any case a discussion of the results is humbled by the lack of information we had at hand about the content expected to be large once applied between the classes.

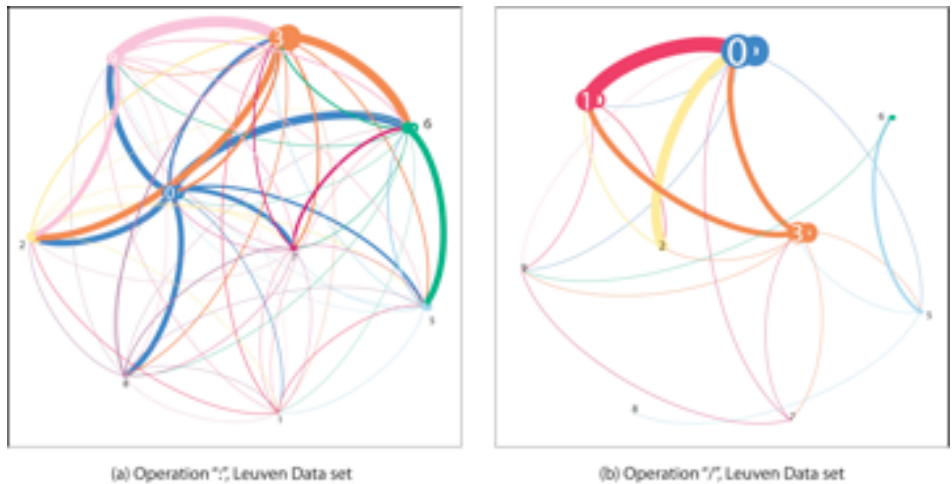

(a) Operation ":", Leuven Data set    (b) Operation "/", Leuven Data set

**Figure 2:** *(a) Network of UDC classes in the Leuven dataset by the use of the operation (a) ":" (Simple relation), (b) "/" (Consecutive extension)*

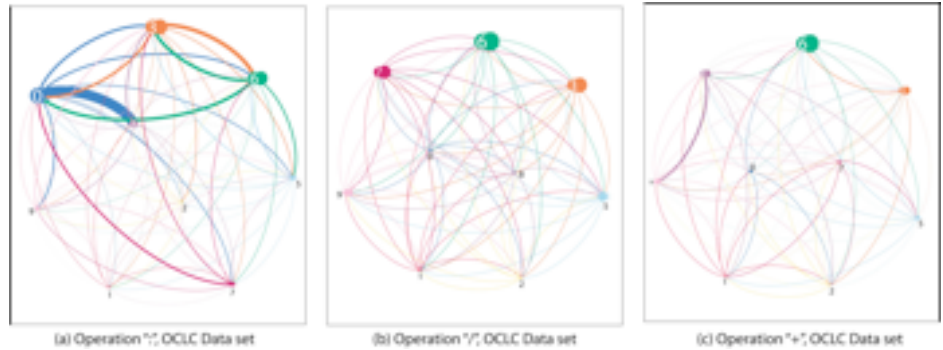

**Figure 4:** *Network of UDC classes in the OCLC dataset by the use of the operation (a) ":" (Simple relation), (b) "/" (Consecutive extension), (c) "+" (Addition)*

## 7. Conclusion

In this paper we present different methods of analysing sets of UDC numbers retrieved from library collections. We argue that quantitative methods and related visualizations can be used to compare different instances of the use of the UDC and also to compare them with the designed system. The UDC represents a complex system of symbolic annotation to codify description of the content of documents. A closer view into compound UDC numbers reveals the complex nature of the UDC itself, which deserves future exploration from a complexity theoretical perspective. The main classes form a hierarchical network. Each auxiliary sign creates additional links between classes. The auxiliaries themselves introduce another kind of nodes into this complex network. At the end the UDC presents itself as a network with different kind of nodes and links. In this paper only a selection of those network-representations are displayed. To raise awareness of the UDC as an object of further study for this kind of research is one aim of this paper.

The interpretation of our results is hampered by the fact that the datasets differ in nature, and are therefore only partly comparable, but also that we have only limited context information for the set of UDC numbers under study. The datasets encompass the Master Reference File (the authority file of the UDC), data on the use of UDC in subject classification in one library (Leuven), and a selection of UDC numbers from various collections (OCLC). In the last case, we have no information from which collection, related to which works, and even more importantly from which period of time the UDC numbers stem. Hence, we focus on the description of the methods.

In a written personal communication the Editor-in-Chief of the UDC Dr. Aida Slavic pointed out, in which respects such an analysis is also of interest to the editors of the UDC. For instance, mapping the size (and growth) of UDC classes might give an indication of future needs for editing and schedule work. On the other hand, statistics about the length of UDC numbers also contains valuable information for the editors. As Dr. Slavic writes:

> "[Currently, we find] [i]n the UDC … over 700 simple numbers that are longer than 11 digits and the longest simple number is 15 digits. Most of them are in class 6. [An example is]:
>
> 621.397.132.129  Other simultaneous systems with single transmission channel
> 621.397.132.122  Frequency multiplex system (FAM)
> 621.397.132.125  PAL (Phase Alternating Line) system
>
> These long notations can indicate two things to editors: a) the subject is very specific (maybe even too specific) and specificity may need reducing and b) the schedule contains enumerated complex subject presented with a simple notation that can be restructured in a faceted way."

The analysis of UDC numbers in collections represents feedback to editors about the use of classes. However, in this case, the temporal provenance of UDC numbers deserves special attention. Across the editions of the UDC, not only are UDC numbers added and deleted, they also are shifted (and re-labeled) and recombined, as well as receiving changed descriptions.

We are convinced that mapping out basic statistics on UDC classes as used in libraries could be of interest both for the information professionals using the UDC in their daily work as well as for users, who might profit from an overview about the nature and focus of a specific collection.

**Acknowledgment**

We would like to thank Ed O'Neill of the OCLC Office of Research who provided us with the OCLC dataset. We would also like to thank Johan Rademakers and Bart Peeters from KU Leuven who provided the Leuven dataset. Aida Slavic gave comments on the paper, and was an indispensable sparring partner for discussion. Part of this work has been funded by the Network of Excellence for Internet Science, FP7 – 288021.

## About authors

**Richard P. Smiraglia** is Professor, Information Organization Research Group, School of Information Studies, at the University of Wisconsin, Milwaukee. He has defined the meaning of "a work" empirically, and has revealed the ubiquitous phenomenon of instantiation among information objects. Recent work includes empirical analysis of social classification, and epistemological analysis of the role of authorship in bibliographic tradition. His "Idea Collider" research team is working on a unified theory of knowledge. An Honorary Fellow of the Virtual Knowledge Studio, Amsterdam, he is a collaborating member of the Knowledge SpaceLab effort to map the evolution of knowledge in Wikipedia. He holds a PhD (1992) from the University of Chicago. He is editor-in-chief of the journal Knowledge Organization.

**Andrea Scharnhorst** is Head of research at Data Archiving and Networked Services (DANS) - an institute of the Royal Netherlands Academy of Arts and Sciences (KNAW). She is also scientific coordinator of the computational humanities programme at the e-humanities group <http://ehumanities.nl <http://ehumanities.nl//) of the KNAW. She has published about the transfer of concepts and methods at the interface between physics, information sciences, social sciences and humanities. Her work focuses on modeling and simulating the emergence of innovations (new modes of behaviour and learning, forms of communication, technologies or scientific ideas) in social systems.

**Almila Akdag Salah** holds a BSc in industrial design, an MA in art history from Istanbul Technical University and a PhD from the Art History Department of UCLA. Her research interests are in the area of technoscience art and its place in the art historical canon; citation networks and mapping of the three semi-related discipline[1]s (cognitive science, visual culture and art history) and their interactions. At the Virtual Knowledge Studio of KNAW she was part of the "Knowledge Space Lab" project, which contributed with a map on Wikipedia and UDC <link to http://www.scimaps.org/maps/map/design_vs_emergence__127/> to the Places and Spaces exhibit. Currently she conducts at the University of Amsterdam, her research project "DeviantArt: Mapping the Alternative Art World", supported by a VENI award from the Netherlands Organization for Scientific Research.

**Cheng Gao** got her B.S in Computer Science from Beijing Language and Culture University (BLCU) (2005) and M.S. in Computer Science from Beijing University of Posts and Telecommunications (BUPT) (2009). She worked as Scientific Programmer for the Knowledge Space Lab project of the Royal Netherlands Academy of Arts and Sciences (KNAW).